\documentclass{article}

\usepackage{arxiv}

\usepackage[utf8]{inputenc} % allow utf-8 input
\usepackage[T1]{fontenc}    % use 8-bit T1 fonts
\usepackage{hyperref}       % hyperlinks
\usepackage{url}            % simple URL typesetting
\usepackage{booktabs}       % professional-quality tables
\usepackage{amsfonts}       % blackboard math symbols
\usepackage{nicefrac}       % compact symbols for 1/2, etc.
\usepackage{microtype}      % microtypography
\usepackage{lipsum}		% Can be removed after putting your text content
\usepackage{graphicx}
\usepackage[numbers,square,sort&compress]{natbib}
\usepackage{doi}
\usepackage{array}
\usepackage{makecell}
\usepackage{amssymb}

\newcolumntype{L}[1]{>{\raggedright\arraybackslash}p{#1}}
\newcolumntype{C}[1]{>{\centering\arraybackslash}p{#1}}

\title{Ghost Traffic: ICMP Tunneling-Based Billing Bypass in LTE Networks}

\author{
  {%\includegraphics[scale=0.06]{orcid.pdf}\hspace{1mm}
  Jung Jin Kim} \\
  78ResearchLab\\
  114, Bongeunsa-ro, Gangnam-gu\\
  Seoul, Republic of Korea, 06123\\
  \texttt{crazyhacker@78researchlab.com} \\
  \And
  {%\includegraphics[scale=0.06]{orcid.pdf}\hspace{1mm}
  Sungyup Nam} \\
  78ResearchLab\\
  114, Bongeunsa-ro, Gangnam-gu\\
  Seoul, Republic of Korea, 06123\\
  \texttt{synam@78researchlab.com} \\
  \And
  \href{https://orcid.org/0000-0002-7116-6062}
  {\includegraphics[scale=0.06]{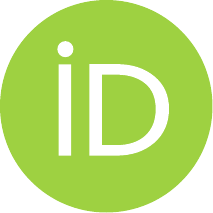}\hspace{1mm}Seungho Jeon}\thanks{Corresponding author.} \\
  Department of Smart Security\\
  Gachon University\\
  1342, Seongnam-daero, Sujeong-gu\\
  Seongnam-si, Gyeonggi-do, Republic of Korea, 13120\\
  \texttt{shjeon90@gachon.ac.kr} \\
}

\date{}

\renewcommand{\shorttitle}{Ghost Traffic: ICMP Tunneling-Based Billing Bypass in LTE Networks}

\hypersetup{
pdftitle={Ghost Traffic: ICMP Tunneling-Based Billing Bypass in LTE Networks},
pdfsubject={Cellular network security},
pdfauthor={JJ Kim, S. Name, and S. Jeon},
pdfkeywords={ICMP tunneling, Billing bypass, Cellular network security, Android, Mobile data charging},
}

\begin{document}
\maketitle

\begin{abstract}
	Cellular data billing is a core operational mechanism for mobile Internet service providers (ISPs), and a policy gap that excludes a specific protocol from usage accounting can lead to a practical security threat. Some cellular ISPs treat ICMP echo traffic as control traffic rather than user data and exclude it from billing. At the same time, Android allows ordinary applications to create ICMP echo sockets without root privileges because of an unsafe default configuration, and the combination of these two conditions forms a vulnerability that can bypass data billing. Existing billing-bypass attacks either require root privileges to create raw sockets and modify routing tables, or do not provide an end-to-end implementation that works in a non-rooted environment, which limits the threat to a small group of experts. This paper proposes Ghost Traffic, an end-to-end system that uses Android's \texttt{VpnService} to encapsulate all application traffic into ICMP echo payloads without root privileges and route it through an external proxy server. The proposed system targets both public IPv4 environments and IPv6-only LTE environments through two variants: IPv4 ICMP tunneling and IPv4-over-IPv6 ICMP tunneling. We evaluated its applicability in seven ISP environments in South Korea, Japan, and the United States, and observed end-to-end tunneling in six of them. We observed that billing bypass occurred in multiple environments and quantitatively showed this effect by measuring that Quality of Service (QoS) throttling was not applied even after the data cap was exhausted. Finally, we propose layered countermeasures across the device, platform, and network levels, performed responsible disclosure, and show that the operational practice of not billing ICMP traffic can lead to practical billing bypass.
\end{abstract}

% keywords can be removed

\section{Introduction}\label{sec:1}
\sloppypar{
Mobile cellular data billing is performed at the Packet Data Network Gateway (P-GW), and the billing system accounts for data usage based on Internet Protocol (IP) packet flows through Charging Data Records (CDRs) \cite{ref-acc,ref-3gpp-32240}. 
Internet Control Message Protocol (ICMP) is a control protocol originally designed for network diagnostics, and the payload field of an echo request message (type=8, code=0) can carry arbitrary data within the Maximum Transmission Unit (MTU) \cite{ref-icmp}. 
Some mobile ISPs maintain an operational practice of treating ICMP traffic as control traffic rather than user data and excluding it from billing policies \cite{ref-no-acc}. 
Meanwhile, Android allows certain process groups to create ICMP echo sockets without root privileges through the kernel parameter \texttt{ping\_group\_range}, which was introduced in Linux 2.6.39 \cite{ref-ping-linux}. 
In Samsung Galaxy series devices, the default value of this parameter is set to \texttt{0 2147483647}, which creates a structural issue in which all ordinary applications can create ICMP echo sockets by calling the \texttt{socket} function.
}
\sloppypar{
The combination of billing exemption for ICMP traffic and the ability to carry arbitrary payloads forms a fundamental security vulnerability \cite{ref-icmp,ref-no-acc,ref-zander-comst07,ref-wendzel-csur15}. 
An attacker can encapsulate Transmission Control Protocol (TCP)/IP packets in ICMP echo payloads and route all application traffic through an external proxy server. 
In environments where these conditions hold, the tunneling traffic may be omitted from usage accounting and lead to billing bypass. Existing billing-bypass techniques, such as TCP retransmission exploits and Voice-over-Long Term Evolution (VoLTE) signaling exploits, require root privileges on the device to create raw sockets and modify routing tables \cite{ref-tcp-ex,ref-volte-ex,ref-volte-ex2}. 
In contrast, the attack proposed in this study is qualitatively different in its threat level because it works with only ordinary Android application privileges. 
Considering the default \texttt{ping\_group\_range} configuration in Samsung Galaxy series devices, this issue can apply to many Android devices that are used worldwide, substantially expanding the scope of potential attackers. 
Moreover, the proxy server required for the attack can be built using free or low-cost cloud services, which effectively eliminates the attacker's cost barrier and may cause large-scale financial damage to cellular ISPs over the long term.
}
\sloppypar{
Prior studies on cellular network billing issues have analyzed several attack vectors, but they share important limitations \cite{ref-pol-vul1,ref-pol-vul2,ref-proto-acc,ref-volte-ex2}. 
Go et al. \cite{ref-tcp-ex} demonstrated a free-riding attack that exploits the policy of some ISPs excluding TCP retransmission packets from billing. 
However, root privileges are required for packet manipulation, and major carriers have already applied partial patches for this issue. 
Li et al. \cite{ref-volte-ex} showed that the VoLTE signaling bearer can deliver arbitrary data without billing and used ICMP tunneling as an attack mechanism, but this approach also assumes a rooted device for raw socket creation and routing table modification. 
Hong et al. \cite{ref-no-acc} systematically analyzed ICMP and TCP billing policies across six ISPs in the United States and South Korea and confirmed that Korean carriers did not bill ICMP echo traffic. 
However, they did not provide a working tunneling system or quantitatively measure the actual amount of bypassed data. 
The common limitations of these studies can be summarized as two issues: dependence on rooted devices and the absence of an end-to-end implementation for non-rooted environments.
}
\sloppypar{
This paper proposes Ghost Traffic, an ICMP tunneling-based billing-bypass system that operates without root privileges by using the Android \texttt{VpnService} API. 
The proposed system creates a local Tunnel (TUN) interface, intercepts all IP traffic generated by applications on the device, encapsulates each packet into the payload of an ICMP echo request, and sends it to an external proxy server. 
The proxy server decapsulates the original packet and forwards it to the destination server using ordinary TCP/IP. 
The implementation consists of two variants: IPv4 ICMP tunneling and IPv4-over-IPv6 ICMP tunneling, which addresses the IPv6-only bearer issue in LTE environments. 
We evaluated the proposed system in seven cellular ISP environments and, to the extent possible, separately verified address allocation, tunnel establishment, performance, and billing bypass for each environment. 
For ethical reasons and responsible disclosure, we anonymize each ISP using a country code and an index. 
We also responsibly disclosed the identified issues to the relevant operators before publication. 
The contributions of this study are summarized as follows.
}
\begin{itemize}
\item We implemented a complete ICMP tunneling-based billing-bypass system that works without root privileges by exploiting the structural interaction between the Android \texttt{VpnService} API and \texttt{ping\_group\_range}.
\item We demonstrated the abuse potential of ICMP non-billing policies in seven cellular ISP environments across South Korea, Japan, and the United States, showing that the vulnerability may not be limited to a specific country or carrier.
\item We quantitatively evaluated the practical usability and performance limitations of the attack channel by measuring throughput, latency, and TCP retransmissions across different payload sizes and ISP environments.
\item We proposed layered countermeasures at the Android, device firmware, and ISP network levels, providing practical defense guidance for both carriers and platform vendors.
\end{itemize}

\sloppypar{
The rest of this paper is organized as follows. 
Section~\ref{sec:2} explains the cellular network billing architecture and the policy gap in ICMP traffic billing. 
Section~\ref{sec:3} reviews related work. 
Section~\ref{sec:4} presents the attack design and implementation of Ghost Traffic. 
Section~\ref{sec:5} evaluates the proposed attack across multiple ISP environments using four research questions. 
Section~\ref{sec:6} presents layered countermeasures, and Section~\ref{sec:7} discusses ethical considerations and responsible disclosure. 
Finally, Section~\ref{sec:8} concludes the paper.
}

\section{Background}\label{sec:2}
\subsection{Cellular Network Accounting Architecture}\label{sec:2.1}

\begin{figure}[t]
\centering
\includegraphics[width=\linewidth]{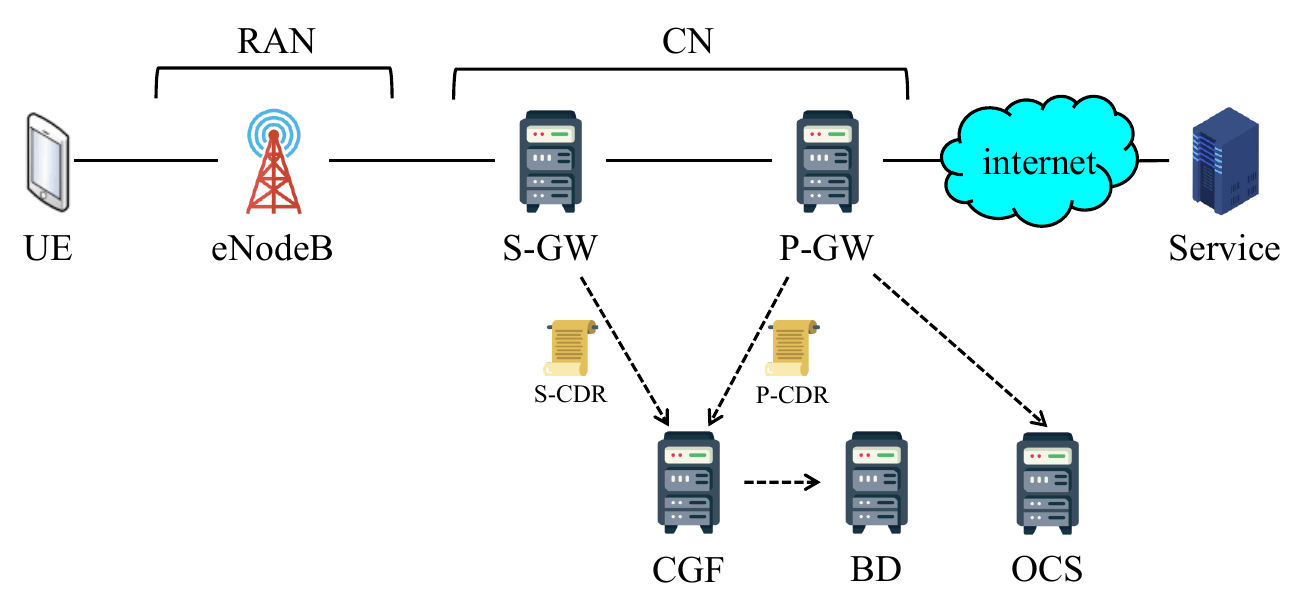}
\caption{Cellular Network Accounting Architecture.}
\label{fig:cellular_accounting_architecture}
\end{figure}

\sloppypar{
An LTE cellular network broadly consists of a Radio Access Network (RAN) and a Core Network (CN) (see Figure~\ref{fig:cellular_accounting_architecture}). 
The RAN handles the radio segment between the User Equipment (UE) and the CN. 
IP packets generated by the UE are delivered through the eNodeB and then forwarded to the Serving Gateway (S-GW) in the CN through a General Packet Radio Service Tunneling Protocol-User Plane (GTP-U) tunnel before being forwarded to the external Internet through the P-GW \cite{ref-lte-arch1,ref-lte-arch2}. 
When the UE starts a data service, it first sends a Packet Data Protocol (PDP) context activation request to the S-GW. 
Through this process, the S-GW and P-GW create CDRs for the session and begin collecting charging information \cite{ref-acc}. 
The S-GW collects radio resource usage, while the P-GW collects external data network usage.
}

\sloppypar{
Cellular data billing is operated through two modes: offline charging and online charging \cite{ref-3gpp-32240}. 
In offline charging, the Charging Trigger Function (CTF) in the P-GW generates a charging event when a data session occurs and passes it to the Charging Data Function (CDF) to construct a CDR. 
The generated CDR is delivered through the Charging Gateway Function (CGF) to the Billing Domain (BD) outside the core network, and the billing system in the BD finally calculates the amount to be charged to the user. 
In online charging, a separate Online Charging System (OCS) authorizes usage in advance, and data service is immediately terminated when the user's remaining credit is exhausted.
}
\sloppypar{
A CDR contains user identifiers, session information, source and destination IP addresses, port numbers, and protocol identifiers, and the billing system accounts for data usage based on this information \cite{ref-acc}. 
Most cellular ISPs bill the entire size of IP packets in bytes, but some apply different policies depending on the protocol type \cite{ref-proto-acc,ref-icmp-pol}. 
More specifically, the billing system observes only the IP layer in the middle of the network path and therefore cannot identify the actual content of the traffic or the state of higher-layer protocols \cite{ref-pol-vul1,ref-pol-vul2}. 
As a result, a billing-exemption policy for a specific protocol can become a vector for billing-bypass attacks. 
This structural characteristic, in which a billing system applies differentiated policies based on protocol identifiers, forms the fundamental prerequisite for the ICMP-based billing-bypass attack studied in this paper.
}

\subsection{Billing Policy Gap in ICMP Traffic Accounting}\label{sec:2.2}
\sloppypar{
ICMP is a network-layer control protocol defined by Request for Comments (RFC) 792 and was designed for error reporting and network diagnostics during IP operation \cite{ref-icmp}. 
An ICMP packet consists of a 20-byte IP header, an 8-byte ICMP header, and an optional payload field \cite{ref-rfc791,ref-icmp}. 
The ICMP header consists of a type field that indicates the message type, a code field that distinguishes detailed behavior, a checksum field for error detection, and identifier and sequence number fields that identify request-response pairs. 
Among several ICMP message types, echo request (type=8, code=0) and echo reply (type=0, code=0) are used to check reachability to a target host and can carry arbitrary data in the payload field within the MTU \cite{ref-icmp,ref-zander-comst07}.
}
\sloppypar{
Because ICMP was originally designed for control purposes rather than user data transfer, some mobile ISPs exclude ICMP echo traffic from billing, and this policy differs across countries and ISPs. 
Hong et al. \cite{ref-icmp-pol} empirically analyzed ICMP billing policies across six major ISPs in South Korea and the United States and found that all three South Korean ISPs did not bill ICMP echo request and reply messages. 
In contrast, all three U.S. ISPs treated ICMP echo traffic as normally billable traffic, showing that ICMP billing-exemption policies differ by ISP. 
This policy inconsistency arises from the gap between the design intent of ICMP as a control protocol and actual network operation practices. 
In ISP environments that apply a billing-exemption policy, it creates the condition under which ICMP payloads can be abused as data tunnels.
}
\sloppypar{
For ICMP echo packets to be forwarded normally in a cellular network, three conditions must be satisfied. 
First, the total packet size, including the payload, must not exceed the standard MTU of 1,500 bytes; 
otherwise, IP fragmentation may occur or packets may be dropped. Second, the identifier value in the ICMP header of the echo reply must match the identifier value in the echo request; 
otherwise, the cellular gateway drops the reply packet. 
Third, the checksum field in the ICMP header must be correctly calculated, and packets with invalid checksums are discarded within the cellular network. 
A notable point is that although RFC 792 requires the payload of an echo reply to be identical to that of the echo request \cite{ref-icmp,ref-rfc1122}, actual cellular environments do not verify payload equality. 
This means that echo request and reply payloads can technically carry different contents and be used as a bidirectional data channel, which is a key prerequisite for the ICMP tunneling attack proposed in this paper.
}

\section{Related Work}\label{sec:3}
\sloppypar{
Studies on cellular billing bypass have focused on how networks treat specific protocols or traffic types in their billing systems. 
Prior work has analyzed policy gaps through several paths, including DNS, TCP retransmission, VoLTE signaling, and ICMP. 
However, ICMP-based end-to-end tunneling systems that operate in non-rooted Android environments have not been sufficiently studied. 
This section reviews prior work in three categories: billing-policy loopholes, TCP retransmission-based bypass, and VoLTE signaling-based bypass.
}

\subsection{Billing Policy Loopholes in Cellular Networks}\label{sec:3.1}
\sloppypar{
Peng et al. \cite{ref-pol-vul1,ref-pol-vul2} discovered a policy gap in 3G cellular networks in which data communication over Domain Name Service (DNS) ports was not billed, as well as techniques for hiding data usage from the billing system through source IP spoofing and Time-to-Live (TTL) manipulation. 
They first demonstrated that free-riding and over-billing attacks are possible in real 3G networks and pointed out the structural limitation that billing systems observe only the IP layer. 
However, most carriers are now aware of these attack vectors and have already responded through policy changes or traffic blocking.
}
\sloppypar{
Hong et al. \cite{ref-icmp-pol} systematically analyzed billing policies for ICMP echo traffic and TCP packets across six ISPs in the United States and South Korea and demonstrated that all three Korean ISPs did not bill ICMP echo messages. 
They further analyzed mechanisms by which some carriers exempt application traffic from billing using only IP addresses and confirmed the possibility of a free-riding attack that disguises traffic accordingly. 
However, their study was limited to policy analysis and conceptual confirmation of attack feasibility, and did not implement a fully working tunneling system or quantitatively evaluate performance and bypassed data volume.
}

\subsection{Billing Bypass via TCP Retransmission}\label{sec:3.2}
\sloppypar{
Go et al. \cite{ref-proto-acc} analyzed TCP retransmission billing policies across 12 ISPs in six countries and found that nine ISPs in the United States, China, and Europe applied a ``blind'' policy that unconditionally billed retransmission packets, whereas three South Korean ISPs used a ``selective'' policy that excluded retransmission packets from billing. 
Based on this observation, they implemented two attacks: a usage-inflation attack that injects malicious retransmissions against blind-policy ISPs to inflate a victim's data usage to a monthly cap within minutes, and a free-riding attack that encapsulates real TCP packets in fake TCP headers disguised as retransmission packets against selective-policy ISPs to obtain 15.6--22.1 Mbps of unbilled bandwidth. 
However, the client implementation of the free-riding attack uses the \texttt{vtund} daemon to directly control a TUN/TAP virtual interface \cite{ref-proto-acc,ref-linux-tuntap}. 
This fundamentally limits practical abuse by ordinary users because it requires rooting or jailbreaking the device.
}

\subsection{Billing Bypass via VoLTE Signaling}\label{sec:3.3}
\sloppypar{
Li et al. \cite{ref-volte-ex} found that access control for the VoLTE signaling bearer is absent in LTE networks and demonstrated that arbitrary data can be transmitted without billing by exploiting the policy that exempts VoLTE signaling traffic from charging. 
More specifically, they implemented an ICMP tunneling attack that encapsulates data packets into ICMP packets through raw sockets and modifies routing tables to send them to the Internet through the VoLTE signaling bearer. 
They confirmed that unbilled data transmission was possible for more than 10 hours with a maximum throughput of 16 Mbps. 
However, this attack is fundamentally limited in ordinary user environments because both raw socket creation and routing table modification are possible only on rooted devices.
}
\sloppypar{
Kim et al. \cite{ref-volte-ex2} discovered four unbilled data channels that abuse VoLTE infrastructure: phone-to-phone direct communication, phone-to-Internet communication, Session Initiation Protocol (SIP) tunneling, and media tunneling. 
They also showed that over-billing attacks are possible through caller phone number spoofing. 
This work is significant because it systematically analyzed VoLTE-related vulnerabilities from both the control-plane and data-plane perspectives and demonstrated them in two major U.S. carriers. 
However, as with Li et al. \cite{ref-volte-ex}, the attack implementation assumes the presence of VoLTE infrastructure and shares the limitation that root privileges are required to transmit traffic to arbitrary destinations.
}

\section{Attack Design}\label{sec:4}
\sloppypar{
This section explains the threat model, preconditions, client implementation, and proxy server implementation of Ghost Traffic. 
The key idea is to capture application traffic using Android \texttt{VpnService} and then encapsulate it into ICMP echo payloads for delivery to an external proxy server. 
We also present two tunneling variants that apply depending on the address-allocation environment and distinguish the applicability of each variant.
}

\subsection{Attack Overview}\label{sec:4.1}

\sloppypar{
The attacker assumed in this paper is an ordinary Android application user without root privileges, holding only the \texttt{INTERNET} and \texttt{BIND\_VPN\_SERVICE} permissions \cite{ref-android-vpn}. 
The attacker does not require additional device modifications or special equipment. 
The attack has two preconditions: the ISP to which the attacker's device belongs must exclude ICMP echo traffic from billing, and the attacker must be able to operate a proxy server in an external network. 
At the time of this study, many free or low-cost cloud servers were available, so the infrastructure cost for the attacker is low.
}

\begin{figure}[t]
    \centering
    \includegraphics[width=\linewidth]{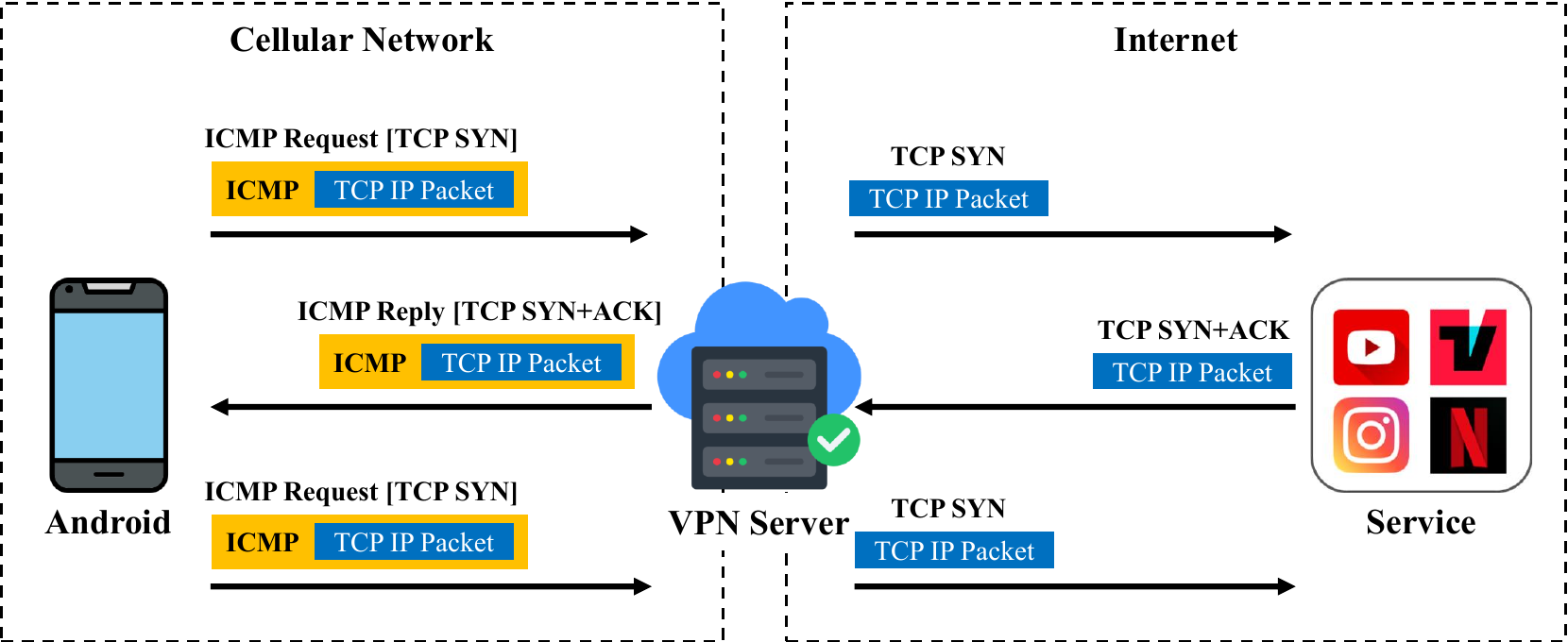}
    \caption{Overview of Ghost Traffic attack.}
    \label{fig:ghost_traffic_overview}
\end{figure}

\sloppypar{
Ghost Traffic consists of three components: an Android client, an ICMP tunnel, and a proxy server. 
Figure \ref{fig:ghost_traffic_overview} shows the overall architecture. 
The Android client captures all IP traffic generated by applications on the device through the \texttt{VpnService} API \cite{ref-android-vpn,ref-android-vpn-imc16}, encapsulates it into ICMP echo payloads, and sends it to the proxy server \cite{ref-icmp,ref-zander-comst07,ref-wendzel-csur15}. The proxy server decapsulates the original packets, forwards them to destination servers using ordinary TCP/IP, and returns response packets to the device by embedding them in ICMP echo replies. During this process, the ISP billing system observes only the ICMP header and cannot recognize the actual data encapsulated inside the payload. Therefore, in ISPs that exclude ICMP echo traffic from billing, user traffic inside the payload may not be accounted for in usage accounting \cite{ref-icmp-pol}.
}

\subsection{Device-side Preconditions}\label{sec:4.2}
\sloppypar{
It has traditionally been understood that directly creating and sending ICMP packets requires raw sockets, and raw socket creation requires root privileges. 
However, the \texttt{ping\_group\_range} kernel parameter introduced in Linux 2.6.39 specifies the group ID range that can create ICMP echo sockets without raw sockets. 
Processes in that range can create IPv4 ICMP echo sockets without root privileges simply by calling \texttt{socket(AF\_INET, SOCK\_DGRAM, IPPROTO\_ICMP)} \cite{ref-ping-linux}. 
Although this parameter is exposed under \texttt{/proc/sys/net/ipv4}, Linux uses the same permission control for IPv6 ping sockets. 
Therefore, processes in the allowed group ID range can also create IPv6 ICMP echo sockets by calling \texttt{socket(AF\_INET6, SOCK\_DGRAM, IPPROTO\_ICMPV6)} \cite{ref-pingv6-linux}. 
At the time of this study, the default value of this parameter in Samsung Galaxy series devices was confirmed to be \texttt{0 2147483647}. 
This value represents the lower and upper bounds of the group IDs allowed to create ICMP echo sockets and means that all group IDs from 0 to 2147483647 are allowed. 
Since Android applications are assigned unique UID/GID values when installed, ordinary application UID/GID values are also included in this allowed range. 
Therefore, ordinary applications that declare only the \texttt{android.permission.INTERNET} permission can create IPv4 and IPv6 ICMP echo sockets without root privileges.
}
\sloppypar{
To capture all application traffic on a device in a non-rooted environment, Ghost Traffic uses the Android \texttt{VpnService} API \cite{ref-android-vpn,ref-android-vpn-imc16}. 
It can be activated simply by declaring the \texttt{BIND\_VPN\_SERVICE} permission and obtaining user consent for a Virtual Private Network (VPN) connection. 
When \texttt{VpnService} is activated, the system creates a local TUN interface and redirects all outbound IP traffic generated by applications on the device to that interface. 
The \texttt{VpnService} application can read and process these packets through the TUN file descriptor and write them back. 
Sockets used by \texttt{VpnService} itself are excluded from the VPN tunnel through the \texttt{protect()} method, preventing a loop in which packets re-enter the TUN interface \cite{ref-android-vpn}.
}

\subsection{Android Client Implementation}\label{sec:4.3}

\begin{figure}[t]
    \centering
    \includegraphics[width=\linewidth]{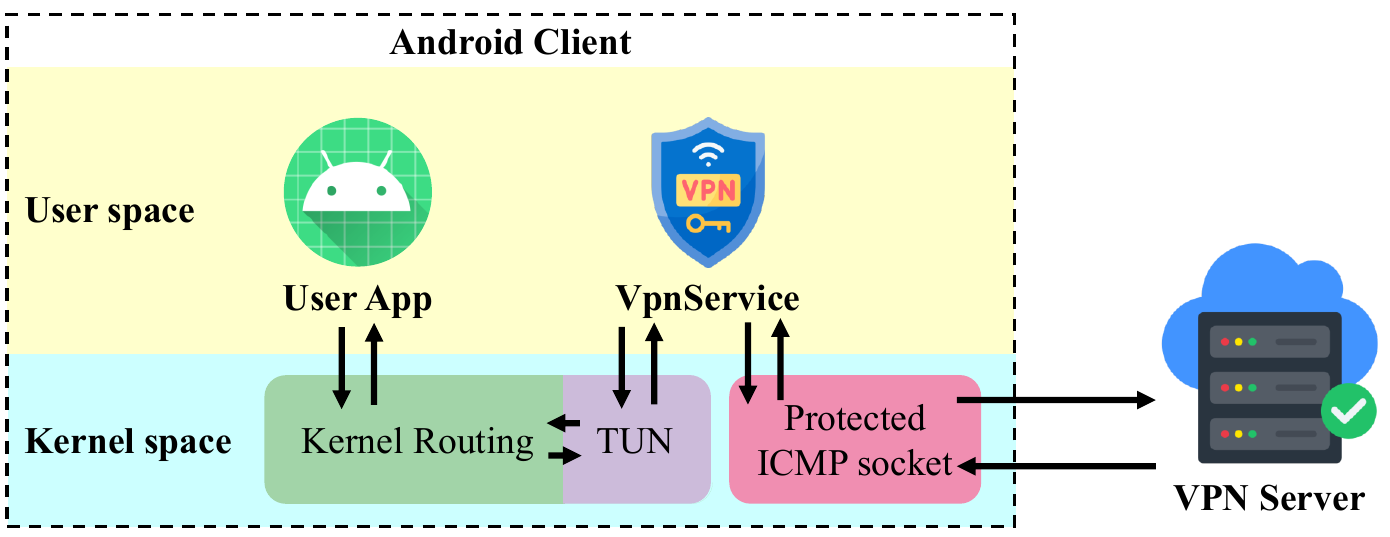}
    \caption{Packet processing in Android client.}
    \label{fig:android_client_processing}
\end{figure}

\sloppypar{
Figure \ref{fig:android_client_processing} shows the Android client implementation of Ghost Traffic. 
The Android client creates a local TUN interface through the \texttt{VpnService} API and reads outbound IP packets generated by applications on the device from this interface.
When \texttt{addRoute("0.0.0.0", 0)} of \texttt{VpnService.Builder} is specified during TUN interface creation, the Android system sets the default route of the routing table to the TUN interface \cite{ref-android-vpn-builder}. 
Subsequently, outbound IP packets generated by each application are automatically delivered to the TUN interface according to the kernel routing table. 
Therefore, the client can obtain original IP packets simply by calling \texttt{read()} on the TUN file descriptor, without separate packet-interception logic. 
Because this process uses Android's ordinary routing mechanism, each application operates as if it were performing normal TCP/IP communication and is unaware that its packets are being delivered through the VPN path. 
As a result, Ghost Traffic can transparently capture all outbound traffic from applications without modifying the implementation of each application. 
Meanwhile, the ICMP socket used by \texttt{VpnService} itself is excluded from the TUN interface through \texttt{protect()}, preventing ICMP packets sent to the proxy server from re-entering the TUN interface.
}
\sloppypar{
Captured IP packets are encapsulated in the payload field of ICMP echo requests and sent to the proxy server \cite{ref-zander-comst07,ref-wendzel-csur15}. The encapsulation process handles three factors. 
First, to satisfy the MTU constraint, if the size of the original packet exceeds the MTU minus the IP header (20 bytes) and the ICMP header (8 bytes), the packet is split and processed \cite{ref-rfc791,ref-icmp,ref-rfc1191}. 
Second, to identify request-response pairs and distinguish multiple sessions, the client assigns a session ID to the identifier field of the ICMP header. 
When generating echo replies, the proxy server sets the same identifier value to satisfy the forwarding condition of the cellular gateway. 
Third, the client correctly calculates and sets the ICMP checksum for packet integrity verification. This step is essential because packets with invalid checksums are dropped in cellular networks \cite{ref-icmp}.
}
\sloppypar{
In the downlink direction, the client receives ICMP echo reply packets sent by the proxy server, extracts the original response packets from the payload, and writes them to the file descriptor of the TUN interface using \texttt{write()} so that they are delivered to the Android network stack. 
The Android system treats packets received from the TUN interface as ordinary network responses. 
Therefore, from the application's perspective, the VPN tunnel is not visible and communication proceeds in the same way as ordinary TCP/IP communication. 
In this way, the client transparently tunnels the traffic of all applications installed on the device, applies tunneling without separate application-level modification or configuration, and can lead to billing bypass in environments where ICMP echo is not billed.
}

\subsection{Proxy Server Implementation}\label{sec:4.4}

\begin{figure}[t]
    \centering
    \includegraphics[width=\linewidth]{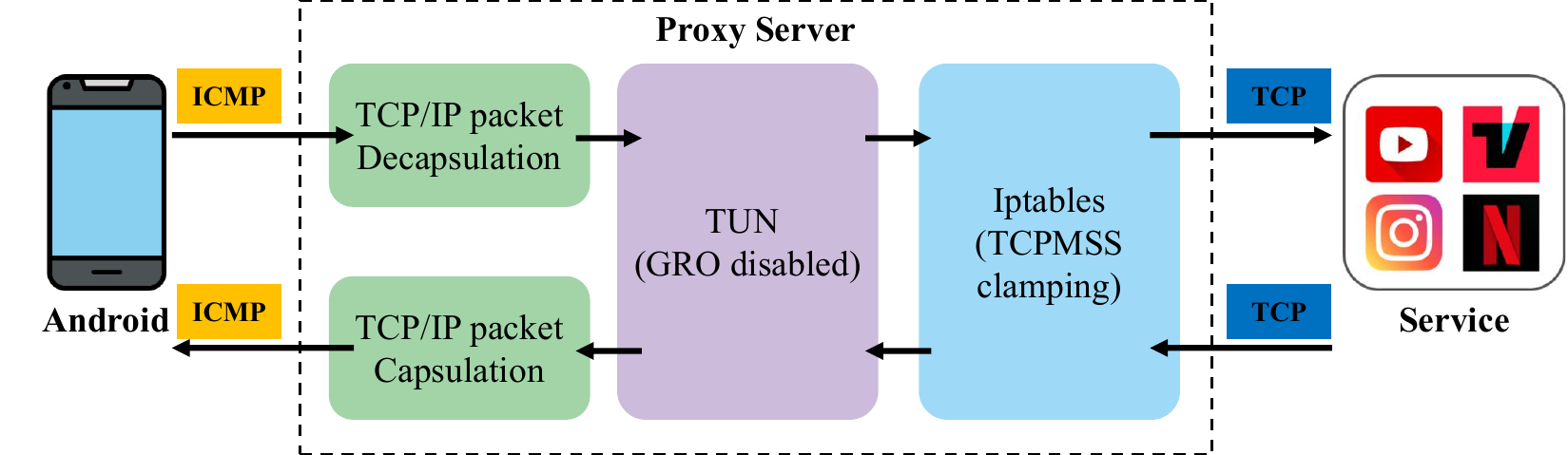}
    \caption{Packet processing in proxy server.}
    \label{fig:proxy_server_processing}
\end{figure}

\sloppypar{
The proxy server extracts the payload from ICMP echo request packets received from the Android client, restores the original IP packets, and writes them to a TUN interface created on the server \cite{ref-linux-tuntap}. 
Packets written to the TUN interface are translated to the public IP address of the proxy server by a MASQUERADE rule in the POSTROUTING chain of \texttt{iptables} and are then forwarded to the external Internet. 
In addition, a TCPMSS clamp rule is applied to adjust the TCP Maximum Segment Size (MSS) value to the Path MTU (PMTU), ensuring that TCP connections operate normally even in the reduced effective MTU environment caused by ICMP tunneling \cite{ref-rfc1191,ref-rfc8201}.
}
\sloppypar{
When response packets from the destination server are received from the external Internet, the proxy server encapsulates them into the payload of ICMP echo replies and returns them to the client. 
In this process, the identifier field of the ICMP header in the echo reply must be set to the same identifier value as that of the previously received echo request; 
otherwise, the cellular gateway drops the reply packet. 
In addition, the proxy server must prevent its kernel from automatically generating replies to ICMP echo requests so that the proxy server program can directly control responses to client requests. 
Because this automatic-response suppression parameter is separated by the address family to which the echo request arrives, \texttt{/proc/sys/net/ipv4/icmp\_echo\_ignore\_all} must be set to 1 for IPv4 ICMP tunnels, and \texttt{/proc/sys/net/ipv6/icmp/echo\_ignore\_all} must be set to 1 for IPv4-over-IPv6 ICMP tunnels \cite{ref-ping-linux}.
}
\sloppypar{
During implementation, a problem may occur because of the Generic Receive Offload (GRO) feature of the server-side Network Interface Card (NIC). 
GRO is a hardware optimization feature that merges multiple small packets into a larger packet to reduce CPU load in high-speed communication environments \cite{ref-linux-offload}. 
As a result, the size of packets received by the proxy server may exceed the MTU set by the client. 
To address this problem, packets should be routed through the TUN interface rather than the physical NIC. 
Since the TUN interface operates with GRO disabled, the kernel should re-segment merged packets according to the MTU size before delivering them.
}

\subsection{Tunnel Variants}\label{sec:4.5}

\sloppypar{
The first variant, IPv4 ICMP tunneling, operates in environments where the ISP assigns a public IPv4 address to the device. 
In this case, both the Android client and the proxy server have IPv4 addresses, so the ICMP echo socket can be created using \texttt{socket(AF\_INET, SOCK\_DGRAM, IPPROTO\_ICMP)}, and the tunnel can be configured without additional network settings by specifying the IPv4 address of the proxy server as the destination. 
A virtual IPv4 address is assigned to the TUN interface created by \texttt{VpnService}, and the source address of captured application traffic is set to this virtual address and translated to the public address through NAT at the proxy server. 
This variant is simple to configure and does not require additional protocol translation, but it cannot be applied in IPv6-only LTE environments.
}
\sloppypar{
The second variant, IPv4-over-IPv6 ICMP tunneling, targets ISPs that assign only IPv6 addresses in LTE environments. 
Some ISPs assign only IPv6 addresses to devices on LTE bearers, so the first IPv4-based variant cannot be directly applied \cite{ref-3gpp-23401}. 
To address this problem, Ghost Traffic adopts a dual structure in which the ICMP tunnel between the client and proxy server is configured over IPv6 using \texttt{socket(AF\_INET6, SOCK\_DGRAM, IPPROTO\_ICMPV6)}, while communication between the proxy server and the Internet is handled over IPv4. 
Since Android \texttt{VpnService} internally operates on IPv4, application traffic is captured as IPv4 packets. 
By encapsulating these packets into IPv6 ICMP echo payloads, IPv4 Internet communication becomes possible even in IPv6-only LTE environments \cite{ref-rfc6877}.
}
\sloppypar{
Both variants assume that a public address is directly assigned to the device. 
If a private IPv4 address is assigned to the device and ICMP echo packets pass through carrier-grade NAT, the public source address of the client observed by the proxy server may change from packet to packet. 
In this case, the proxy server cannot determine the destination address to which echo replies should be returned, so the tunnel cannot be established. 
The same problem also occurs when a device has an IPv6 address but sends IPv4 ICMP rather than IPv6 ICMP, because the carrier translates the packet into IPv4 and the translated source IPv4 address can continuously change. 
Therefore, IPv4 ICMP tunneling requires a public IPv4 address to be directly assigned to the device, while IPv4-over-IPv6 ICMP tunneling requires a public IPv6 address to be directly assigned to the device.
}

\begin{table}[t]
\centering
\caption{Comparison of ICMP tunneling approaches.}
\label{tab:tunnel_variants}
\renewcommand{\arraystretch}{1.15}
\resizebox{\textwidth}{!}{%
\begin{tabular}{
L{0.28\textwidth}
!{\vrule width 1.2pt}
L{0.31\textwidth}
|L{0.31\textwidth}
}
\Xhline{1.2pt}
\textbf{} &
\textbf{IPv4 ICMP Tunneling} &
\textbf{IPv4-over-IPv6 ICMP Tunneling} \\
\Xhline{1.2pt}
\textbf{Target Environment} &
ISP with IPv4 allocation &
IPv6-only LTE ISP \\
\hline
\textbf{ICMP Socket} &
\texttt{AF\_INET} &
\texttt{AF\_INET6} \\
\hline
\textbf{Client-Proxy Tunnel} &
IPv4 ICMP &
IPv6 ICMP \\
\hline
\textbf{Proxy-Internet Communication} &
IPv4 &
IPv4 \\
\hline
\textbf{NAT Required} &
Yes &
No \\
\Xhline{1.2pt}
\end{tabular}}
\end{table}

\sloppypar{
The two variants differ only in the tunnel protocol between the client and the proxy server, while the proxy server's decapsulation and forwarding logic is identical. 
Table \ref{tab:tunnel_variants} summarizes the target environments and main characteristics of the two variants. 
IPv4 ICMP tunneling is simple and has low overhead in environments where IPv4 addresses are assigned. 
In contrast, IPv4-over-IPv6 ICMP tunneling can be applied to IPv6-only environments and has the advantage of enabling direct client-proxy communication without Network Address Translation (NAT) because public IPv6 addresses are directly assigned to devices in LTE. 
In practice, the appropriate variant should be selected by checking the type of IP address assigned to the device.
}

\section{Evaluation}\label{sec:5}
\sloppypar{
This section evaluates the effectiveness of Ghost Traffic from multiple perspectives using four Research Questions (RQs).
}

\begin{itemize}
\item \textbf{RQ1.} Does Ghost Traffic normally tunnel representative application traffic in real service environments?
\item \textbf{RQ2.} Is the performance overhead caused by ICMP tunneling low enough for practical use?
\item \textbf{RQ3.} How does the applicability of Ghost Traffic differ depending on ISP-specific address allocation and ICMP handling policies?
\item \textbf{RQ4.} Does actual billing bypass occur in ISP environments that apply a billing-exemption policy?
\end{itemize}

\subsection{Experimental Setup}\label{sec:5.1}

\begin{table}[t]
\centering
\caption{Anonymized ISP environments used in our evaluation.}
\label{tab:isp_envs}
\renewcommand{\arraystretch}{1.15}
\resizebox{\textwidth}{!}{%
\begin{tabular}{
L{0.18\textwidth}
|L{0.12\textwidth}
|L{0.22\textwidth}
|L{0.42\textwidth}
}
\Xhline{1.2pt}
\textbf{Country} &
\textbf{ISP ID} &
\textbf{Network Type} &
\textbf{Evaluation Scope} \\
\Xhline{1.2pt}
South Korea & KR-A & Local cellular network &
Functional, billing, and performance tests \\
\hline
South Korea & KR-B & Local cellular network &
Functional, billing, and performance tests \\
\hline
South Korea & KR-C & Local cellular network &
Functional, billing, and partial performance tests \\
\hline
Japan & JP-A & Local cellular network &
Functional and partial performance tests \\
\hline
Japan & JP-B & Local cellular network &
Functional, billing, and partial performance tests \\
\hline
Japan & JP-C & Local cellular network &
Address-allocation test \\
\hline
United States & US-A & Local cellular network &
Functional and performance tests \\
\Xhline{1.2pt}
\end{tabular}}
\end{table}

\sloppypar{
Table \ref{tab:isp_envs} summarizes the seven cellular ISP environments used in this study. 
Considering responsible disclosure and the possibility of identifying operators, each ISP is denoted using an anonymized identifier that combines a country code and an index. 
The Android client created a local TUN interface using the \texttt{VpnService} API and encapsulated IP packets generated by the device into ICMP echo payloads. 
The proxy server was deployed on a Google Cloud Platform (GCP)-based Linux server, and restored packets were forwarded to the external Internet using a server-side TUN interface, \texttt{iptables} NAT, and a TCPMSS clamp rule. 
In IPv6-only LTE environments, the tunnel between the client and proxy server was configured over IPv6 ICMP, while Internet communication after the proxy server was handled over IPv4. 
We could not perform the same experiments for all ISPs because some ISPs did not assign public IPv6 addresses at the time of the experiment, some changed their ICMP handling policies after responsible disclosure, and measurement tools were limited for specific USIM and device combinations.
}
\sloppypar{
In the evaluation scope, a functional test means an experiment that checks whether Ghost Traffic can tunnel IP packets from a real application or traffic generator end-to-end. 
A billing test means an experiment that checks whether ICMP tunneling bypasses the billing policy by observing data usage or QoS throttling in an actual cellular plan environment. 
A performance test means an experiment that compares throughput, Round-Trip Time (RTT), and TCP retransmissions between ordinary TCP/IP communication and ICMP tunneling communication. 
An address-allocation test means an experiment that checks whether the device is assigned an IPv4 or IPv6 address and determines the applicability of IPv4 ICMP tunneling or IPv4-over-IPv6 ICMP tunneling. 
Throughput was measured using \texttt{iperf3} \cite{ref-iperf3}, and the ICMP payload size was set to 500B, 1000B, and 1500B. 
Each throughput experiment was performed for 60 seconds and repeated five times under the same configuration. 
RTT was measured using \texttt{ping} by sending 1,000 packets at 10 ms intervals for each payload size. 
Long-term transfer stability was measured by running \texttt{iperf3} for 1,000 seconds at a 10 Mbps rate and comparing receiver throughput and the number of TCP retransmissions.
}

\subsection{RQ1: Functional Verification}\label{sec:5.2}
\sloppypar{
To evaluate RQ1, we incrementally verified whether the proposed system tunnels traffic end-to-end regardless of the transport-layer protocol and application type. 
First, we checked whether ICMP echo sockets could be created without root privileges on the device side to verify the socket-creation feasibility required by the attack. 
Next, we checked the address type assigned to the device in each ISP environment and determined the applicable tunnel variant based on that type. 
In environments where the tunnel could be configured, we verified end-to-end delivery using two traffic types. 
The first was ICMP reachability, where we used \texttt{ping} to check whether echo requests and replies could make a round trip through the tunnel. 
The second was TCP application traffic, where we used large transfers with \texttt{iperf3} \cite{ref-iperf3} and real applications performing video streaming and file downloads. 
At each step, normal operation was defined as the process in which IP packets captured on the device are encapsulated into ICMP echo payloads, transmitted to the proxy server, decapsulated and forwarded to the destination, and then restored through the same path. 
We also performed verification without modifying applications, checking whether applications operated as in ordinary TCP/IP communication without recognizing the presence of the VPN tunnel.
}

\begin{table}[t]
\centering
\caption{Functional verification results across ISP environments.}
\label{tab:functional_verification}
\renewcommand{\arraystretch}{1.15}
\resizebox{\textwidth}{!}{%
\begin{tabular}{
L{0.08\textwidth}
|L{0.15\textwidth}
|L{0.15\textwidth}
|L{0.22\textwidth}
|C{0.12\textwidth}
|C{0.13\textwidth}
|C{0.15\textwidth}
}
\Xhline{1.2pt}
\textbf{ISP} &
\textbf{Network Type} &
\textbf{Allocated Address} &
\textbf{Tunnel Variant} &
\textbf{ICMP Echo Socket} &
\textbf{ICMP Reachability} &
\textbf{TCP Application Traffic} \\
\Xhline{1.2pt}
KR-A & Local cellular & Public IPv6 & IPv4-over-IPv6 ICMP &
\checkmark & \checkmark & \checkmark \\
\hline
KR-B & Local cellular & Public IPv6 & IPv4-over-IPv6 ICMP &
\checkmark & \checkmark & \checkmark \\
\hline
KR-C & Local cellular & Public IPv6 & IPv4-over-IPv6 ICMP &
\checkmark & \checkmark & \checkmark \\
\hline
JP-A & Local cellular & Public IPv6 & IPv4-over-IPv6 ICMP &
\checkmark & \checkmark & \checkmark \\
\hline
JP-B & Local cellular & Public IPv6 & IPv4-over-IPv6 ICMP &
\checkmark & \checkmark & \checkmark \\
\hline
JP-C & Local cellular & No public IPv6 & -- &
\checkmark & -- & -- \\
\hline
US-A & Local cellular & Public IPv6 & IPv4-over-IPv6 ICMP &
\checkmark & -- & \checkmark \\
\Xhline{1.2pt}
\end{tabular}}
\end{table}

\sloppypar{
Table \ref{tab:functional_verification} summarizes the functional verification results for the seven ISP environments. 
Device-side ICMP echo socket creation succeeded in all seven environments regardless of ISP, because socket creation is determined by the device's \texttt{ping\_group\_range} configuration rather than by the network. 
In the six environments except JP-C, public IPv6 addresses were assigned to the device, so the IPv4-over-IPv6 ICMP tunneling variant was applied. 
Among them, ICMP reachability and TCP application traffic were both delivered normally in KR-A, KR-B, KR-C, JP-A, and JP-B, and applications also operated transparently. 
In US-A, TCP application traffic was delivered end-to-end through the ICMP tunnel, confirming functional operation and indicating that the applied variant worked correctly. 
However, because the prepared rooted device did not recognize the USIM, we conducted the experiment using a personal device. 
Since high-frequency \texttt{ping} transmission requires root privileges, an independent ICMP reachability measurement was not performed.
}
\sloppypar{
In the JP-C environment, a public IPv6 address was not assigned to the device, so the IPv4-over-IPv6 variant could not be applied. 
Therefore, ICMP reachability and TCP application traffic verification were not performed. 
Nevertheless, device-side ICMP echo socket creation was confirmed normally, showing that the cause of non-applicability was not the device but the network's address-allocation policy. 
The relationship between address-allocation type and applicable variant is analyzed in detail in Section~\ref{sec:5.4}. 
Overall, ICMP reachability and TCP application traffic were normally tunneled in all six environments where end-to-end functional verification was possible. 
This shows that the proposed system does not depend on a specific protocol or application. 
In addition, the fact that all verification was performed without modifying applications means that an ordinary user could apply billing bypass simply by installing an application.
}

\subsection{RQ2: Performance Evaluation}\label{sec:5.3}
\sloppypar{
To evaluate RQ2, we measured three metrics: throughput, latency, and long-term stability. 
All measurements used ordinary TCP/IP communication as the baseline and compared it with ICMP tunneling, while varying the payload size among 500, 1000, and 1500B. The target environments were ISP environments in which end-to-end tunneling worked. 
JP-C was excluded from performance measurements because no public IPv6 address was assigned and the tunnel could not be configured. 
Throughput and stability were measured using \texttt{iperf3} \cite{ref-iperf3}, and latency was measured using \texttt{ping}. 
In some cells, receiver reports were missing or the distinction between baseline and tunneling was unclear, so reliable values could not be derived. 
These cells are marked in the table with the reason. Throughput and stability report both sender and receiver values to reveal losses inside the tunnel, and throughput overhead was calculated based on receiver goodput as perceived by the user.
}

\begin{table}[t]
\centering
\caption{Throughput test for 60s per run.}
\label{tab:throughput}
\renewcommand{\arraystretch}{1.15}
\resizebox{\textwidth}{!}{%
\begin{tabular}{
L{0.09\textwidth}
|L{0.16\textwidth}
|L{0.32\textwidth}
|L{0.32\textwidth}
}
\Xhline{1.2pt}
\textbf{ISP} &
\textbf{Payload (Bytes)} &
\textbf{Baseline Sender/Receiver (Mbps)} &
\textbf{Tunneling Sender/Receiver (Mbps)} \\
\Xhline{1.2pt}
KR-A & 500 &
$29.02 \pm 7.39$ / $28.16 \pm 7.50$ &
$0.199 \pm 0.007$ / $0.064 \pm 0.007$ \\
\hline
KR-A & 1000 &
$8.34 \pm 0.55$ / $7.31 \pm 0.59$ &
$0.201 \pm 0.003$ / $0.066 \pm 0.002$ \\
\hline
KR-A & 1500 &
$9.29 \pm 0.67$ / $8.53 \pm 0.57$ &
$0.203 \pm 0.003$ / $0.067 \pm 0.003$ \\
\hline
KR-B & 500 &
$17.08 \pm 1.28$ / $16.80 \pm 1.27$ &
$8.27 \pm 1.93$ / $8.02 \pm 1.88$ \\
\hline
KR-B & 1000 &
$13.84 \pm 0.79$ / $13.62 \pm 0.79$ &
$10.84 \pm 0.21$ / $10.60 \pm 0.17$ \\
\hline
KR-B & 1500 &
$18.52 \pm 3.79$ / $18.24 \pm 3.81$ &
$10.53 \pm 1.85$ / $10.36 \pm 1.90$ \\
\hline
KR-C & 500/1000/1500 &
$54.32 \pm 1.47$ / $54.10 \pm 1.46$ &
Not measured \\
\hline
JP-A & 500/1000/1500 &
Not reliably measured &
Not reliably measured \\
\hline
JP-B & 500/1000/1500 &
Not reliably measured &
Not reliably measured \\
\hline
JP-C & -- &
Not applicable &
Not applicable \\
\hline
US-A & 1000 &
$8.49 \pm 0.82$ / $7.79 \pm 0.59$ &
$4.49 \pm 2.21$ / $3.94 \pm 2.08$ \\
\Xhline{1.2pt}
\end{tabular}}
\vspace{0.5em}

\footnotesize{
Note: KR-C baseline was measured at approximately 53--55 Mbps
for all three payload sizes. KR-C tunneling was not measured because
public IPv6 allocation was no longer available after responsible disclosure.
JP-C was not applicable because no public IPv6 address was allocated.
}
\end{table}

\begin{figure}[t]
    \centering
    \includegraphics[width=0.7\linewidth]{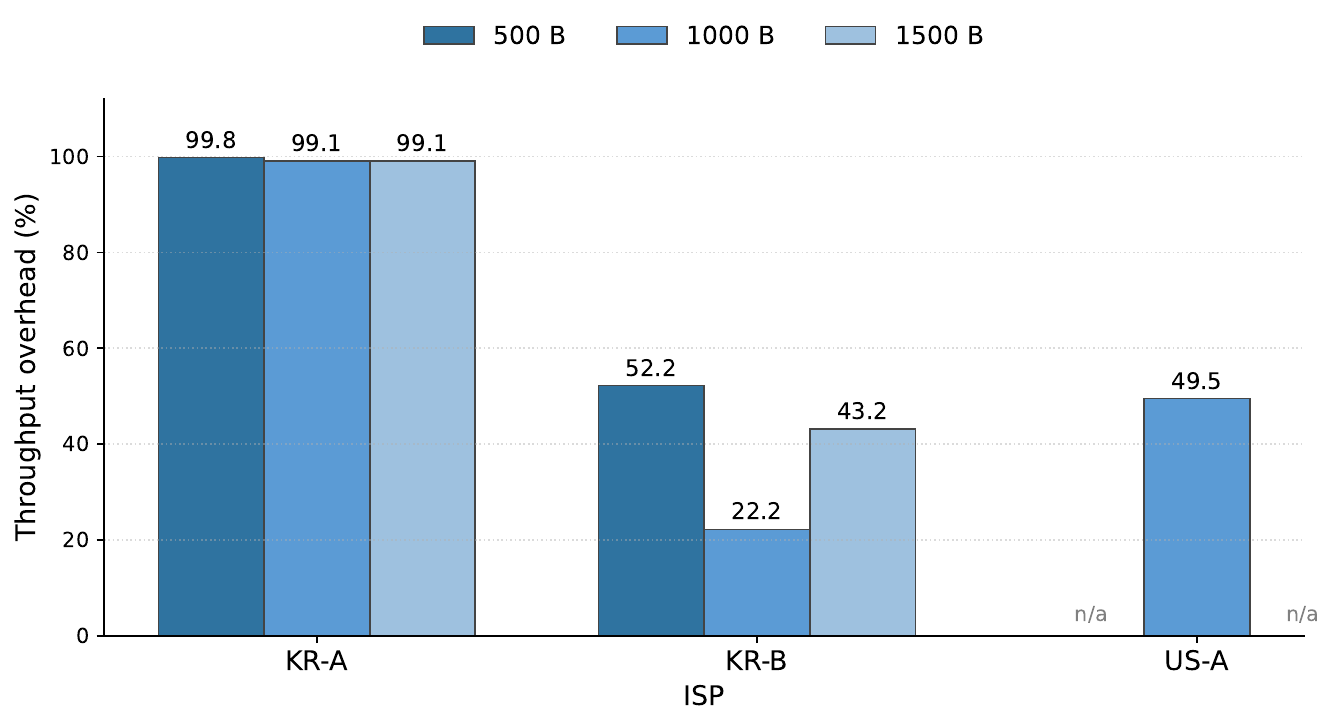}
    \caption{Throughput overhead according to the payload size.}
    \label{fig:throughput_overhead}
\end{figure}

\sloppypar{
Throughput was measured for 60 seconds with five repetitions, and the mean and standard deviation were calculated after excluding failed runs that ended with a receiver report of zero. 
Table \ref{tab:throughput} summarizes baseline and tunneling throughput and overhead for each ISP and payload, and Figure \ref{fig:throughput_overhead} shows the overhead as a bar graph. 
For KR-A, KR-B, and US-A, the baseline and tunneling measurements were clearly distinguished and values were derived. 
For KR-C, public IPv6 was no longer assigned after responsible disclosure, so tunneling throughput could not be measured and only the baseline remained. 
For JP-A, most runs ended with a receiver value of zero, and for JP-B, thousands of retransmissions were recorded even in runs that appeared to be the baseline. 
Therefore, reliable throughput values could not be derived for these two environments. For US-A, only the 1000B payload was stably distinguished.
}

\sloppypar{
Table \ref{tab:throughput} and Figure \ref{fig:throughput_overhead} show that tunneling throughput differs greatly by ISP. 
KR-A showed approximately 99\% overhead for all payload sizes, and tunneling throughput dropped to about 0.07 Mbps. 
However, the KR-A result does not represent a general characteristic of the ISP; 
rather, it reflects a large drop in tunneling speed caused by KR-A applying strong QoS to ICMP traffic after responsible disclosure. 
In contrast, KR-B showed overhead between 22\% and 52\% and maintained 8--11 Mbps even under tunneling, while US-A showed about 50\% overhead at 1000 bytes. 
The large difference between KR-A and KR-B in the same country is due to ISP-specific handling of ICMP echo traffic and policy changes after responsible disclosure. 
No consistent monotonic relationship appeared between payload size and overhead, and in KR-B the overhead was lowest at 1000 bytes. The addition of a 20-byte IP header and an 8-byte ICMP header to each packet and the fragmentation of large packets are basic overhead factors, but the large measured variation shows that the network's ICMP handling policy is the more dominant factor. 
Differences in ISP-specific handling policies are analyzed in Section~\ref{sec:5.4}.
}

\begin{table}[t]
\centering
\caption{Round-trip time for 1000 pings at 10ms interval.}
\label{tab:rtt}
\renewcommand{\arraystretch}{1.15}
\resizebox{\textwidth}{!}{%
\begin{tabular}{
L{0.11\textwidth}
|L{0.18\textwidth}
|L{0.30\textwidth}
|L{0.30\textwidth}
}
\Xhline{1.2pt}
\textbf{ISP} &
\textbf{Payload (Bytes)} &
\textbf{Baseline RTT} &
\textbf{Tunneling RTT} \\
\Xhline{1.2pt}
KR-A & 500 &
$73.74 \pm 100.95$ &
$618.63 \pm 283.98$ \\
\hline
KR-A & 1000 &
$43.73 \pm 19.24$ &
$609.51 \pm 262.06$ \\
\hline
KR-A & 1500 &
No reply &
$397.40 \pm 133.99$ \\
\hline
KR-B & 500 &
$31.74 \pm 10.26$ &
$546.98 \pm 276.82$ \\
\hline
KR-B & 1000 &
$29.80 \pm 5.06$ &
$488.90 \pm 260.91$ \\
\hline
KR-B & 1500 &
$32.70 \pm 6.04$ &
$310.41 \pm 193.69$ \\
\hline
KR-C & 500 &
$39.83 \pm 8.80$ &
Not measured \\
\hline
KR-C & 1000 &
$39.41 \pm 6.87$ &
Not measured \\
\hline
KR-C & 1500 &
$42.28 \pm 9.62$ &
Not measured \\
\hline
JP-A & 500 &
$100.82 \pm 47.29$ &
$12696.57 \pm 6703.91$ \\
\hline
JP-A & 1000 &
$130.25 \pm 114.13$ &
$17101.81 \pm 6650.20$ \\
\hline
JP-A & 1500 &
$128.66 \pm 85.14$ &
$15453.43 \pm 6936.04$ \\
\hline
JP-B & 500 &
$150.59 \pm 64.75$ &
$722.92 \pm 265.01$ \\
\hline
JP-B & 1000 &
$168.39 \pm 57.46$ &
$708.25 \pm 266.02$ \\
\hline
JP-B & 1500 &
$144.41 \pm 55.57$ &
$599.61 \pm 272.03$ \\
\hline
US-A & 500/1000/1500 &
Not measured &
Not measured \\
\Xhline{1.2pt}
\end{tabular}}
\vspace{0.5em}

\footnotesize{
Note: The KR-A 1500B baseline did not receive replies because
the gateway did not return echo replies for packets exceeding 1500B.
US-A was not measured because the prepared rooted device did not
recognize the USIM and high-frequency ping required root privileges.
}
\end{table}

\sloppypar{
Latency was measured by sending 1,000 \texttt{ping} packets at 10 ms intervals for each payload size. 
Table \ref{tab:rtt} summarizes the average RTT values for the baseline and tunneling cases with standard deviations, where the standard deviation corresponds to the \texttt{mdev} value reported by \texttt{ping} and thus indicates jitter. In the measured environments, baseline RTT was on the order of tens of milliseconds. 
Baseline and tunneling RTT were both measured in KR-A, KR-B, JP-A, and JP-B, while tunneling RTT was not measured in KR-C because public IPv6 was not assigned. 
US-A was not measured because the rooted device did not recognize the USIM and high-frequency \texttt{ping} required root privileges. 
The 1500-byte baseline of KR-A was not measured because the gateway did not return replies for echo packets exceeding 1500 bytes, and the policy reason for this is discussed in Section~\ref{sec:5.4}.
}
\sloppypar{
Tunneling substantially increased RTT in all measured environments. 
In KR-A and KR-B, tunneling RTT increased to 300--620 ms, roughly ten times the baseline, and jitter also increased. In JP-B, tunneling RTT was 600--720 ms. 
In JP-A, the average tunneling RTT reached 12,000--17,000 ms, which was noticeably worse than in other environments and consistent with the severe loss observed in the throughput measurement for the same environment. 
The main causes of RTT increase are the additional path through the proxy server and retransmission waiting caused by lost packets. 
The increased jitter indicates that the tunnel could not maintain stable delay. 
As payload size increased, tunneling RTT tended to decrease slightly in KR-A and KR-B, but the standard deviations were large, so this trend cannot be concluded with confidence.
}

\begin{table}[t]
\centering
\caption{Sustained-transfer stability over 1000s.}
\label{tab:stability}
\renewcommand{\arraystretch}{1.15}
\resizebox{\textwidth}{!}{%
\begin{tabular}{
L{0.07\textwidth}
|L{0.10\textwidth}
|L{0.22\textwidth}
|L{0.18\textwidth}
|L{0.22\textwidth}
|L{0.18\textwidth}
}
\Xhline{1.2pt}
\textbf{ISP} &
\textbf{Payload (Bytes)} &
\textbf{Baseline Sender/Receiver (Mbps)} &
\textbf{Baseline Retransmissions} &
\textbf{Tunneling Sender/Receiver (Mbps)} &
\textbf{Tunneling Retransmissions} \\
\Xhline{1.2pt}
KR-A & 500 & 7.43 / 7.37 & 0 & 0.074 / 0.066 & 122 \\
\hline
KR-A & 1000 & 10.0 / 10.0 & 30 & 0.075 / 0.066 & 144 \\
\hline
KR-A & 1500 & 10.0 / 10.0 & 3 & 0.074 / 0.066 & 127 \\
\hline
KR-B & 500 & 10.0 / 10.0 & 0 & 10.0 / 9.99 & 2992 \\
\hline
KR-B & 1000 & 10.0 / 10.0 & 16 & 10.0 / 9.99 & 2665 \\
\hline
KR-B & 1500 & 10.0 / 10.0 & 0 & 10.0 / 9.99 & 775 \\
\hline
KR-C & 500 & 10.0 / 10.0 & 0 & Not measured & -- \\
\hline
KR-C & 1000 & 10.0 / 10.0 & 31 & Not measured & -- \\
\hline
KR-C & 1500 & 10.0 / 10.0 & 0 & Not measured & -- \\
\hline
JP-A & 500 & 1.44 / 1.42 & 0 & 0.247 / 0.00 & 2675 \\
\hline
JP-A & 1000 & 1.00 / 0.98 & 0 & 0.054 / 0.00 & 6 \\
\hline
JP-A & 1500 & 1.01 / 0.99 & 0 & 0.735 / 0.699 & 5180 \\
\hline
JP-B & 500 & 10.0 / 0.00 & 0 & 6.20 / 6.19 & 8666 \\
\hline
JP-B & 1000 & 10.0 / 0.00 & 0 & 3.67 / 0.00 & 9029 \\
\hline
JP-B & 1500 & 10.0 / 0.00 & 0 & 3.55 / 3.52 & 7725 \\
\hline
US-A & 500 & 5.49 / 0.00 & 110 & 2.80 / 2.78 & 7466 \\
\hline
US-A & 1000 & 7.12 / 0.00 & 85 & 7.08 / 7.05 & 2819 \\
\hline
US-A & 1500 & 7.73 / 0.00 & 95 & 4.87 / 4.85 & 3810 \\
\Xhline{1.2pt}
\end{tabular}}
\vspace{0.5em}

\footnotesize{
Note: The JP-B and US-A baselines reached the target rate at the sender,
but \texttt{iperf3} did not report receiver summaries.
KR-C tunneling was not measured because public IPv6 was not assigned.
}
\end{table}

\sloppypar{
Long-term stability was measured by transmitting for 1,000 seconds with a target rate of 10 Mbps and measuring receiver throughput and TCP retransmissions. 
Table \ref{tab:stability} summarizes sender and receiver throughput and retransmission counts for the baseline and tunneling cases. 
Each cell was measured using a single run rather than an average, and sender and receiver values are reported together to show that tunnel transmission may not be maintained normally. 
In the baselines for JP-B and US-A, the sender reached the target rate, but \texttt{iperf3} did not report the receiver summary, so the receiver value was recorded as zero. 
The JP-A baseline remained around 1 Mbps because of constraints in the network itself. KR-C tunneling was not measured because no public IPv6 address was assigned.
}
\sloppypar{
Long-term tunneling stability also differed greatly by ISP. KR-B maintained about 10 Mbps even under tunneling, but retransmissions increased substantially to 775--2992. 
In KR-A, tunneling throughput dropped to about 0.07 Mbps and was effectively difficult to use. 
JP-A and JP-B failed to maintain long-term transfer for some payload sizes because receiver throughput ended at zero. 
US-A received 2.8--7.1 Mbps under tunneling, but retransmissions reached 2800--7500. 
In all environments, tunneling increased retransmissions by hundreds to thousands of times compared with the baseline, showing that ICMP payload-based transmission is vulnerable to loss. 
As in the throughput measurements, the difference between KR-A and KR-B was prominent, confirming that long-term stability is also strongly affected by ISP-specific ICMP handling.
}
\sloppypar{
Overall, the practicality of ICMP tunneling differs sharply depending on the ISP environment. 
KR-B and US-A showed practical throughput of several to around ten Mbps with latency on the order of hundreds of milliseconds. 
In contrast, KR-A dropped to 0.07 Mbps and JP-A showed RTTs of tens of seconds, severely limiting practicality. 
Tunneling substantially increased latency and retransmissions in all environments, but some environments provided enough performance for video streaming and file downloads. 
The fundamental causes of these performance differences, namely ISP-specific address allocation and ICMP handling policies, are analyzed in Section~\ref{sec:5.4}. 
Therefore, the answer to RQ2 is that although the overhead is substantial, practical throughput was observed in some environments, particularly KR-B and some US-A conditions, whereas performance constraints were severe in KR-A and JP-A.
}

\subsection{RQ3: Applicability Across ISP Environments}\label{sec:5.4}
\sloppypar{
To evaluate RQ3, we analyze the applicability of Ghost Traffic using two determining factors. 
The first factor is the address type assigned to the device, which determines the applicable tunnel variant. 
The second factor is the ICMP handling policy of the network, which determines whether the tunnel itself can be established. 
Among the two variants, IPv4 ICMP tunneling works in environments where a public IPv4 address is assigned, and IPv4-over-IPv6 ICMP tunneling works in environments where a public IPv6 address is assigned. 
Therefore, the address type assigned to the device must first be checked before selecting the appropriate variant. 
This section summarizes applicability based on the observed results for these two factors across seven ISP environments.
}

\begin{table}[t]
\centering
\caption{Applicability determinants across ISP environments.}
\label{tab:applicability}
\renewcommand{\arraystretch}{1.15}
\resizebox{\textwidth}{!}{%
\begin{tabular}{
L{0.10\textwidth}
|L{0.18\textwidth}
|L{0.25\textwidth}
|L{0.40\textwidth}
}
\Xhline{1.2pt}
\textbf{ISP} &
\textbf{Allocated Address} &
\textbf{Applicable Variant} &
\textbf{ICMP Echo Handling} \\
\Xhline{1.2pt}
KR-A & Public IPv6 & IPv4-over-IPv6 ICMP &
Arbitrary payload forwarded, but $>$1500B dropped \\
\hline
KR-B & Public IPv6 & IPv4-over-IPv6 ICMP &
Arbitrary payload forwarded \\
\hline
KR-C & Public IPv6 & IPv4-over-IPv6 ICMP &
Arbitrary payload forwarded \\
\hline
JP-A & Public IPv6 & IPv4-over-IPv6 ICMP &
Arbitrary payload forwarded \\
\hline
JP-B & Public IPv6 & IPv4-over-IPv6 ICMP &
Arbitrary payload forwarded \\
\hline
JP-C & No public IPv6 & Not applicable &
Not evaluated \\
\hline
US-A & Public IPv6 & IPv4-over-IPv6 ICMP &
Arbitrary payload forwarded \\
\Xhline{1.2pt}
\end{tabular}}
\end{table}

\sloppypar{
Applicability analysis was performed in two steps: address-allocation checking and ICMP echo forwarding observation. 
First, we checked whether the address assigned to the device in each ISP environment was public IPv4 or public IPv6 and determined the applicable variant. 
Next, we checked whether ICMP echo messages carrying arbitrary payloads were normally forwarded in each environment. 
During this process, we also observed ISP-specific behavior such as drops depending on payload size. 
The identifier-matching and checksum conditions for ICMP echo are common to all environments, as explained in Sections~\ref{sec:2.2} and~\ref{sec:4.3}. 
Therefore, this section focuses on the parts that differ by ISP. 
Table \ref{tab:applicability} summarizes applicability according to these two factors.
}

\sloppypar{
As shown in Table \ref{tab:applicability}, six of the seven environments assigned public IPv6 addresses to the device, so the IPv4-over-IPv6 ICMP tunneling variant was applied. 
Since most ISPs in LTE environments directly assign public IPv6 addresses to devices, this variant is essential in IPv6-only bearer environments. 
Because public IPv6 is directly assigned, the client and proxy server can communicate directly without NAT. 
In contrast, JP-C did not receive a public IPv6 address, so the IPv4-over-IPv6 variant could not be applied, and only address-allocation checking was performed in that environment. 
This shows that applicability directly depends on the network's address-allocation policy. 
Therefore, in practical deployment, the address type assigned to the device must be checked before selecting a variant. 
The fact that the six observed environments span South Korea, Japan, and the United States shows that the practice of assigning public IPv6 addresses is not limited to a specific country.
}
\sloppypar{
Regarding ICMP handling policies, all six environments in which tunneling was possible normally forwarded ICMP echo messages carrying arbitrary payloads. 
This shows that the practice of forwarding ICMP echo traffic as control traffic rather than user data commonly exists across multiple networks. 
However, in KR-A, a specific behavior was observed in which the gateway did not return replies to echo packets when the total packet size including the payload exceeded 1500 bytes. 
Even in this case, the tunnel itself was established normally because the client split and encapsulated the original packets within 1500 bytes. 
No such payload-size-based constraint was observed in the other environments. 
Overall, ICMP handling policy is a condition that determines whether the tunnel can be established, but in the observed environments this condition was generally satisfied and did not greatly limit applicability.
}
\sloppypar{
As a result, the applicability of Ghost Traffic is not limited to a specific country or carrier and was broadly observed in environments in South Korea, Japan, and the United States. 
However, applicability is determined by two conditions. 
First, the address type assigned to the device determines the applicable variant, and the IPv4-over-IPv6 variant cannot be applied in environments that do not receive a public IPv6 address. 
Second, the network must forward ICMP echo messages with arbitrary payloads for the tunnel to be established. 
Since the two variants are designed for public IPv4 environments and IPv6-only environments, respectively, the appropriate variant can be selected according to the address type assigned to the device. 
Therefore, the answer to RQ3 is that applicability is determined by address allocation and ICMP handling policy, and it holds in many environments where both conditions are satisfied.
}

\subsection{RQ4: Billing Bypass Effectiveness}\label{sec:5.5}
\sloppypar{
To evaluate RQ4, we verified whether tunneling traffic is actually excluded from usage accounting in ISP environments that do not bill ICMP echo. 
Billing bypass can be confirmed using two pieces of evidence. The first is that data transmitted through the tunnel is not reflected in the carrier's usage accounting. 
The second is that QoS throttling is not applied even after all plan data is exhausted. 
The second piece of evidence holds only when the billing system does not recognize tunneling traffic as data usage, so it is direct evidence of billing bypass. 
This section checks billing-bypass results for each ISP based on these two pieces of evidence.
}

\begin{table}[t]
\centering
\caption{Billing bypass results across ISP environments.}
\label{tab:billing_bypass}
\renewcommand{\arraystretch}{1.15}
\resizebox{\textwidth}{!}{%
\begin{tabular}{
L{0.10\textwidth}
|L{0.22\textwidth}
|L{0.22\textwidth}
|L{0.38\textwidth}
}
\Xhline{1.2pt}
\textbf{ISP} &
\textbf{Billing Test} &
\textbf{Bypass Result} &
\textbf{Evidence} \\
\Xhline{1.2pt}
KR-A & Performed & Bypassed &
Usage not counted \\
\hline
KR-B & Performed & Bypassed &
Usage not counted \\
\hline
KR-C & Performed & Bypassed &
Usage not counted \\
\hline
JP-A & Not performed & -- &
-- \\
\hline
JP-B & Performed & Bypassed &
Usage not counted, QoS not applied after data cap \\
\hline
JP-C & Not performed & -- &
-- \\
\hline
US-A & Not performed & -- &
-- \\
\Xhline{1.2pt}
\end{tabular}}
\end{table}

\sloppypar{
Table \ref{tab:billing_bypass} shows that billing bypass was observed in all four environments where billing verification was performed. 
In KR-A, KR-B, KR-C, and JP-B, data transmitted through the tunnel was not reflected in the carrier's usage accounting. 
This shows that the policy of treating ICMP echo as control traffic and excluding it from billing can be practically abused in a bypass attack. 
In JP-B, QoS non-application was additionally confirmed along with usage non-accounting, providing the strongest form of evidence. 
JP-A and US-A were not included in the billing verification scope, so no results are presented for them. 
In JP-C, billing bypass itself was not possible because no public IPv6 address was assigned and the tunnel could not be established.
}

\begin{figure}[t]
    \centering
    \includegraphics[width=0.7\linewidth]{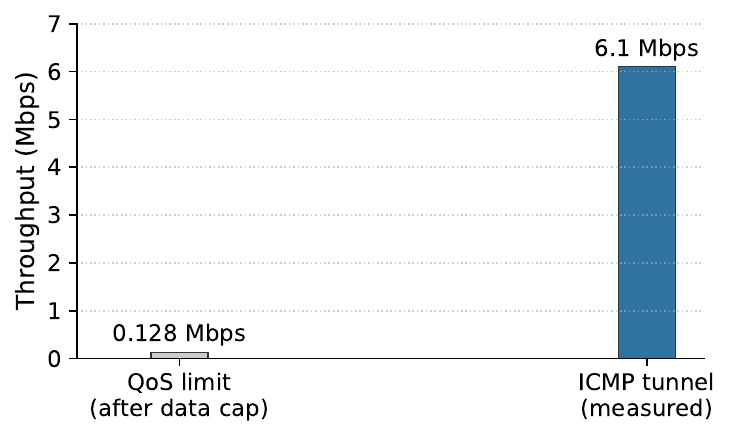}
    \caption{QoS bypass of JP-B after data-cap exhaustion.}
    \label{fig:qos_bypass}
\end{figure}

\sloppypar{
The QoS-bypass measurement performed in the JP-B environment quantitatively shows billing bypass. 
The plan applied a policy that limits the speed to 128 Kbps when all data is exhausted. 
After exhausting all 12 GB of data and entering the QoS-throttled state, the measured speed through the tunnel was about 6.1 Mbps. 
Figure \ref{fig:qos_bypass} compares the QoS throttling rate and the measured rate. 
The measured speed exceeded the QoS limit by about 48 times, directly showing that tunneling traffic was not accounted for as usage. 
If the tunneling traffic had been accounted for, QoS throttling would have been applied and the measured speed would not have exceeded 128 Kbps.
}

\sloppypar{
The identified billing bypass was blocked in the evaluated South Korean ISPs after responsible disclosure. 
KR-A applied strong QoS to ICMP traffic, so the attack became less effective even if the tunnel was established. 
KR-C no longer assigned public IPv6 addresses to the devices we had, making tunnel configuration impossible. 
KR-B changed ICMP packets to be billable, so even though the tunnel could still be established, the traffic was accounted for as usage. 
As a result, at the time of the experiment, billing bypass no longer held in all three South Korean ISPs. This limited post-disclosure performance measurements in some environments. 
The detailed responsible disclosure process is discussed in Section~\ref{sec:7}.
}
\sloppypar{
In summary, Ghost Traffic achieved billing bypass in ISP environments that did not bill ICMP echo. In all four environments where billing verification was performed, tunneling traffic was not reflected in usage accounting. 
In particular, the JP-B QoS-bypass measurement quantitatively demonstrated billing bypass by showing unrestricted speed after data exhaustion. 
After responsible disclosure, the evaluated South Korean ISPs blocked the bypass through patches or billing-policy changes. This shows that the billing-exemption policy itself is the root cause of the bypass. 
Therefore, the answer to RQ4 is that billing bypass can actually occur in environments that maintain an ICMP non-billing policy, and changing that policy blocks the bypass.
}

\subsection{Comparative Study}\label{sec:5.6}
\sloppypar{
This section qualitatively compares Ghost Traffic with prior studies on cellular billing bypass and summarizes the distinctions of this work. 
The comparison targets the five studies discussed in Section~\ref{sec:3}: Peng et al. \cite{ref-pol-vul1,ref-pol-vul2}, Hong et al. \cite{ref-icmp-pol}, Go et al. \cite{ref-proto-acc}, Li et al. \cite{ref-volte-ex}, and Kim et al. \cite{ref-volte-ex2}. 
The comparison uses six metrics: attack vector, target environment, root-privilege requirement, infrastructure dependency, end-to-end implementation, and unbilled data throughput. 
These metrics are key factors that determine the practicality and threat level of an attack. 
Table \ref{tab:comparison} summarizes the five prior studies and Ghost Traffic according to these six metrics.
}

\sloppypar{
The six metrics used in the comparison are defined as follows. 
The attack vector indicates the mechanism abused for billing bypass, and the target environment indicates the network generation and conditions under which the attack holds. 
The root-privilege requirement indicates whether device root privileges are needed to perform the attack and is a key metric that determines the possibility of abuse by ordinary users. 
Infrastructure dependency indicates whether the attack depends on a specific network feature such as VoLTE. 
End-to-end implementation distinguishes whether the study stopped at policy analysis or implemented a complete working system. 
Unbilled data throughput indicates the maximum unbilled data transmission rate reported by each study.
}

\begin{table}[t]
\centering
\caption{Comparative study with prior billing-bypass studies.}
\label{tab:comparison}
\renewcommand{\arraystretch}{1.15}
\resizebox{\textwidth}{!}{%
\begin{tabular}{
L{0.17\textwidth}
|L{0.22\textwidth}
|L{0.13\textwidth}
|L{0.12\textwidth}
|L{0.18\textwidth}
|L{0.12\textwidth}
|L{0.15\textwidth}
}
\Xhline{1.2pt}
\textbf{Reference} &
\textbf{Attack Vector} &
\textbf{Target Environment} &
\textbf{Root Required} &
\textbf{Infrastructure Dependency} &
\textbf{End-to-End System} &
\textbf{Free-Data Throughput} \\
\Xhline{1.2pt}
Peng et al. \cite{ref-pol-vul1,ref-pol-vul2} &
DNS-based charging loophole &
Cellular &
Not reported &
None &
Yes &
Not reported \\
\hline
Hong et al. \cite{ref-icmp-pol} &
ICMP/TCP charging-policy &
Cellular &
Not applicable &
None &
No &
Not reported \\
\hline
Go et al. \cite{ref-proto-acc} &
Spurious TCP-retransmission tunneling &
Cellular &
Yes &
None &
Yes &
15.6--221 Mbps \\
\hline
Li et al. \cite{ref-volte-ex} &
ICMP tunneling over VoLTE signaling bearer &
4G LTE &
Yes &
VoLTE signaling bearer &
Yes &
Up to 16 Mbps \\
\hline
Kim et al. \cite{ref-volte-ex2} &
VoLTE hidden data channels &
4G LTE &
Yes &
VoLTE infrastructure &
Yes &
Up to 21.55 Mbps \\
\hline
Ghost Traffic &
ICMP echo tunneling via \texttt{VpnService} &
Cellular &
No &
Proxy server &
Yes &
Up to 10 Mbps \\
\Xhline{1.2pt}
\end{tabular}}
\end{table}

\sloppypar{
The most important feature of Ghost Traffic in Table \ref{tab:comparison} is that it does not require root privileges. 
The attacks by Go et al. \cite{ref-proto-acc}, Li et al. \cite{ref-volte-ex}, and Kim et al. \cite{ref-volte-ex2} require root privileges on the device to create raw sockets or modify routing tables. 
In contrast, Ghost Traffic works with only ordinary Android application privileges, so non-rooted ordinary users can abuse it directly. 
This difference qualitatively changes the threat level because it expands the attacker population to ordinary users as a whole. 
In addition, the attacks by Li et al. \cite{ref-volte-ex} and Kim et al. \cite{ref-volte-ex2} assume the existence of VoLTE infrastructure. 
Ghost Traffic only requires an external proxy server and does not depend on specific network functions, so its applicability conditions are simpler.
}
\sloppypar{
There is also a difference in implementation completeness. 
Hong et al. \cite{ref-icmp-pol} analyzed ICMP non-billing policies and conceptually confirmed bypass feasibility, but did not implement a working tunneling system. 
In contrast, Ghost Traffic implements a complete end-to-end system including a client and proxy server and quantitatively evaluates it across multiple ISPs. 
In terms of target environment, prior studies are limited to environments that assume 3G or VoLTE, whereas Ghost Traffic was demonstrated in LTE environments where public IPv6 addresses are assigned through the IPv4-over-IPv6 variant. Regarding the reported effect, VoLTE-based studies showed high throughput of 16--21 Mbps, but this was obtained under strong assumptions such as VoLTE infrastructure and root privileges. 
The throughput of Ghost Traffic differs by ISP, but its meaning is different because it is achieved using only ordinary application privileges without those assumptions.
}
\sloppypar{
Overall, Ghost Traffic is distinguished from prior work by simultaneously having four properties: no root-privilege requirement, no dependency on specific infrastructure, an end-to-end implementation, and support for IPv6-only LTE environments. 
Prior studies have individual strengths, but they are bound by prerequisites such as root privileges or VoLTE infrastructure, or they do not reach the level of a working system implementation. 
Ghost Traffic implements a complete system that works in ordinary user environments without these prerequisites. 
In particular, the fact that it operates with ordinary application privileges expands the attacker population and increases the practicality of the threat. 
This is why the threat presented in this study is qualitatively different from prior work. 
Concrete countermeasures for carriers and platform vendors are discussed in Section~\ref{sec:6}.
}

\section{Countermeasures}\label{sec:6}
\sloppypar{
Ghost Traffic becomes possible through the combination of three factors across layers: ICMP echo socket creation on the device, traffic interception by the platform, and the network's practice of not billing ICMP. 
Therefore, defense at a single layer is insufficient, and layered defenses across the device, platform, and network are required. 
At the device and platform layers, the first step is to block rootless ICMP echo socket creation, which is the starting point of the attack. 
This can be achieved by restricting \texttt{ping\_group\_range} in device firmware to system application ranges or disabling it, so that ordinary applications cannot create ICMP echo sockets \cite{ref-ping-linux}. 
Next, the Android system can compare the request and reply payloads of ICMP echo and drop replies that are not identical \cite{ref-icmp,ref-rfc1122}. 
This uses the property of RFC 792 \cite{ref-icmp} that echo and reply data should be identical, so a tunneling channel carrying different data cannot pass this validation. 
However, normal applications that use \texttt{ping} for diagnostic purposes may be affected, and applying firmware updates to many devices may take time.
}
\sloppypar{
At the network layer, carriers can apply defenses that are directly under their control. 
The most direct response is to bill ICMP echo as user data, thereby removing the economic incentive for bypass \cite{ref-icmp-pol,ref-proto-acc}. 
The result in Section~\ref{sec:5.5}, where one ISP blocked the bypass by changing ICMP packets to billable traffic, shows the effectiveness of this approach. 
In addition to billing, carriers can set timeouts for ICMP echo and reply forwarding and detect abnormally large amounts of ICMP echo traffic to rate-limit it \cite{ref-zander-comst07,ref-wendzel-csur15}. 
These network-level measures have the advantage that carriers can apply them without cooperation from devices. 
However, the load on network equipment and the possibility of false positives remain limitations when inspecting all ICMP echo traffic.
}
\sloppypar{
In conclusion, no single layer can completely block the attack alone, so a layered defense combining the device, platform, and network is necessary. 
Converting ICMP traffic to billable traffic at the network level is an immediately applicable and decisive response, and its effect was observed in Section~\ref{sec:5.5}. 
In particular, because the attack operates with ordinary application privileges, network-level blocking can immediately protect the broadest range of users. 
In contrast, firmware and platform responses address the root cause of rootless socket creation, but applying updates to many devices takes time. 
Therefore, carriers should immediately block the bypass through billing-policy changes, while device manufacturers and platform vendors should gradually apply updates that remove the root cause. 
This layered approach improves the overall security level by ensuring that bypass at one layer is blocked at another layer.
}

\section{Ethical Considerations and Responsible Disclosure}\label{sec:7}
\sloppypar{
The purpose of this study is to reveal a real and exploitable billing-bypass vulnerability so that carriers can fix it. 
To prevent harm to third parties during the research process, all experiments were performed only with USIMs and data plans registered under the researchers' own names. 
Therefore, any bypassed data usage was limited to the researchers' own plans, and the experiment scale was restricted to the scope necessary for research to minimize harm to third parties and practical impact on carriers. 
We also did not release tools or code that could reproduce the attack, and we anonymized ISP identifiers throughout the paper to protect carriers until the vulnerability was fully fixed. 
These measures are intended to prevent abuse of the research results while academically conveying the existence and severity of the vulnerability. 
This study was conducted to strengthen defenses and does not encourage unauthorized use.
}
\sloppypar{
Before publication, the researchers responsibly disclosed the vulnerability to affected South Korean ISPs. 
The disclosure was made to one carrier, KR-A. Afterward, all three ISPs applied countermeasures. 
As a result, some measurements could not be completed before the carriers removed the conditions required for the attack, and the missing data in Sections~\ref{sec:5.2} and~\ref{sec:5.3} are due to this circumstance. 
For example, in KR-C, devices that previously received public IPv6 addresses no longer received them from some point, so within the set of devices available to us the tunnel could no longer be configured and tunneling performance could not be measured. 
The detailed patches applied by each carrier and their billing effects are discussed in Section~\ref{sec:5.5}.
}
\sloppypar{
In the South Korean ISPs to which disclosure was made, billing bypass no longer holds because of the applied countermeasures. 
However, the vulnerability disclosure was limited to South Korean carriers, so environments in other countries where disclosure has not been performed may still be affected. 
For this reason, we maintain anonymization throughout the paper to protect carriers that may not yet be protected. We recognize the need for additional responsible disclosure for the remaining environments. 
We recommend that carriers that may not yet be protected apply the layered countermeasures presented in Section~\ref{sec:6}. 
Through this, we hope that the vulnerability revealed by this study leads to stronger defenses rather than attacks.
}

\section{Conclusion}\label{sec:8}
\sloppypar{
This paper proposed Ghost Traffic, an attack that bypasses LTE data billing without root privileges. 
Ghost Traffic intercepts application traffic using Android's \texttt{VpnService}, encapsulates it into ICMP echo packets, and communicates with the external Internet through a proxy server. 
Through two variants, IPv4 ICMP tunneling and IPv4-over-IPv6 ICMP tunneling, it targets both public IPv4 environments and IPv6-only environments. 
In the evaluation, we verified Ghost Traffic mainly using IPv4-over-IPv6 ICMP tunneling and observed functional operation in multiple ISP environments across South Korea, Japan, and the United States. 
Performance differed greatly by ISP, but in some environments it was practical enough to use, and applicability was determined by address allocation and ICMP handling policies. 
We also observed that billing bypass occurred in real environments and quantitatively demonstrated it by measuring that QoS throttling was not applied even after data was exhausted.
}
\sloppypar{
Unlike prior studies that require root privileges or VoLTE infrastructure, Ghost Traffic operates with only ordinary Android application privileges. 
This expands the target of billing-bypass threats from a small group of experts to ordinary users as a whole, making the threat identified in this study qualitatively different from prior work. 
This paper proposed layered countermeasures across the device, platform, and network. 
Among them, converting ICMP traffic to billable traffic at the network level is the most decisive measure, and its effect was observed through responsible disclosure. 
In conclusion, this study shows that the operational practice of treating ICMP as unbilled control traffic can lead to practical billing bypass. 
Carriers should revisit such billing-exemption policies to block similar bypasses.
}

\bibliographystyle{unsrtnat}
\bibliography{references}  %%% Uncomment this line and comment out the ``thebibliography'' section below to use the external .bib file (using bibtex) .





% \end{thebibliography}

\end{document}